\documentclass[journal]{IEEEtran}
\usepackage{lineno}
\usepackage{amsmath,amssymb}
\usepackage{float}
\newtheorem{prop}{Proposition}
\newtheorem{lemma}{Lemma}

\usepackage{makecell}
\usepackage{CJK} 
\usepackage{float}
\usepackage{multirow}
\usepackage{graphicx}
\usepackage{multirow}
\usepackage{tabu,bm}
\usepackage{color}
\usepackage{subfigure}
\usepackage{booktabs}

\usepackage[linesnumbered,ruled,vlined]{algorithm2e}
\usepackage{algorithmic}
\usepackage{url}

\usepackage{cite}

\ifCLASSINFOpdf
\else
\fi

\begin{document}
%
\title{Semantic Communication with Memory}
%
%
%
\author{Huiqiang Xie,~\IEEEmembership{Graduate Student Member,~IEEE,} Zhijin~Qin, \IEEEmembership{Senior~Member,~IEEE,} \\ and 
Geoffrey Ye Li,~\IEEEmembership{Fellow,~IEEE}
\thanks{Huiqiang Xie is with the School of Electronic Engineering and Computer Science, Queen Mary University of London, London E1 4NS, UK (e-mail: h.xie@qmul.ac.uk).  }
\thanks{Zhijin Qin is with the Department of Electronic Engineering, Tsinghua University, Beijing, China (e-mail: qinzhijin@tsinghua.edu.cn).  } 
\thanks{Geoffrey Ye Li is with School of Electrical and Electronic Engineering, Imperial College London, London SW7 2AZ, UK (e-mail: geoffrey.li@imperial.ac.uk).
}
}

\maketitle

\begin{abstract}
While semantic communication succeeds in \textcolor{black}{efficiently} transmitting due to the strong capability to extract the essential semantic information, it is still far from the intelligent or human-like communications. In this paper, we introduce an essential component, memory, into semantic communications to mimic human communications. Particularly, we investigate a deep learning (DL) based semantic communication system with memory, named Mem-DeepSC, by considering the scenario question answer task. We exploit the universal Transformer based transceiver to extract the semantic information and introduce the memory module to process the context information. Moreover, we derive \textcolor{black}{the relationship between the length of semantic signal and the channel noise} to validate the possibility of dynamic transmission. Specially, we propose two dynamic transmission methods to enhance the transmission reliability as well as to reduce the communication overheads by masking some unessential elements, which are recognized through training the model with mutual information.  Numerical results show that the proposed Mem-DeepSC is superior to  benchmarks in terms of answer accuracy and transmission efficiency, i.e., number of transmitted symbols.


\end{abstract}


%
\IEEEpeerreviewmaketitle


\section{Introduction}

The seamlessly connected world fosters unique services, like virtual reality (VR), mobile immersive eXtended reality (XR), or autonomous driving, and brings new challenges to communication systems, such as the scarcity of resources, the congestion of network traffic, and the scalable connectivity for edge intelligence \cite{LetaiefSLL22}. To materialize the vision, semantic communication \cite{qin2021semantic} is a communication paradigm by directly delivering the meanings of information,  and extracting and transmitting only important information relevant to the task at the receiver. In the past couple of years, semantic communication is attracting extensive attention from both academia \cite{qin2021semantic} and industry \cite{tong2022nine, HoydisAVV21}. The latest works take advantage of deep learning (DL) to design end-to-end semantic communication systems for various types of source reconstruction  and specific tasks execution.  Semantic communication has shown a great potential to increase the reliability in performing intelligent tasks, reducing the network traffic, and thus alleviating spectrum shortage. 

The existing works in semantic communication can be categorized by the types of source data. For image-based semantic communication systems, Jankowski \textit{et al.} \cite{JankowskiGM21} have developed digital and analog deep joint source-channel coding (JSCC) to perform the person/car re-identification task directly, which improves the image retrieval accuracy effectively. Lee \textit{et al.} \cite{LeeLCC19} have considered the image classification as the communication task, where JSCC is based on the convolutional neural network (CNN). Moreover, Hu \textit{et al.} \cite{qiyu22} have designed robust semantic communication against semantic noise by employing adversarial training, which reduces the probability of misleading in classification. In order to reduce the communication overheads, Yang \textit{et al.} \cite{yang21} have developed bandwidth-limited semantic communication by removing the redundancy of semantic features while keeping similar classification accuracy. Shao \textit{et al.}~\cite{ShaoMZ22} have proposed a dynamic semantic communication system to adaptively adjust the  number of the active semantic features under different signal-to-noise ratios (SNRs) with a graceful classification accuracy degradation. Bourtsoulatze~\textit{et al.}~\cite{BourtsoulatzeKG19} have investigated the deep image transmission semantic communication systems, in which the semantic and channel coding are optimized jointly. Kurka~\textit{et al.} \cite{KurkaG20} extended Bourtsoulatze's work with the channel feedback to improve the quality of image reconstruction. Huang~\textit{et al.} \cite{HuangTGL21} have designed the image semantic coding method by introducing the framework of rate-distortion, which can save the number of bits as well as keep the good quality of the reconstructed image.

Apart from image based semantic communication systems, the video based semantic communication systems also attracted much attention. Tung \textit{et al.} \cite{Tung21} have designed the initial deep video semantic communications by accounting for occlusion/disocclusion and camera movements. Especially, the authors considered the DL-based frame design for the video reconstruction. Wang \textit{et al.} \cite{wang2022wireless} have proposed the adaptive deep video semantic communication systems by learning to allocate the limited channel bandwidth within and among video frames to maximize the overall transmission performance. Jiang \textit{et al.} \cite{jiang2022wireless} have investigated the application of semantic communications in the video conference, in which the proposed system can maintain high resolution by transmitting some keypoints to represent motions and keep the low communication overheads. Similarly, Tandon \textit{et al.} \cite{Pulkit2021} also considered the video conference transmission. Different from  \cite{jiang2022wireless}, the authors have designed the video semantic communication by converting the video to text at the transmitter and recovering the video from the text at the receiver. By considering the multi-user scenario for multi-modal data transmission, Xie \textit{et al.}~\cite{xie2021task} have proposed a unified Transformer based semantic communication framework to support the image and text transmission and to enable the receiver performing various multimodal tasks. 

Meanwhile, there exist works on the speech-based, text-based, and multimodal semantic communication systems.  Weng \textit{et al.} \cite{WengQL21} have developed the speech recognition-oriented semantic communication, named DeepSC-SR, in which the transmitter sends the speech signal and the receiver restores the text directly. Han \textit{et al.} \cite{22han} have designed an more energy-efficient speech-to-text system by introducing the redundancy removal module to lower the transmitted data size. With the depth exploration in semantic communications, Xie~\textit{et al.}~\cite{XieQ21} have developed a powerful joint semantic-channel coding, named DeepSC, to encode text information into various lengths over complex channels. Moreover, Xie~\textit{et al.}~\cite{XieQLJ21} also have proposed an environment-friendly semantic communication system, named L-DeepSC, for capacity-limited devices. \textcolor{black}{Zhou \textit{et al.} \cite{zhou2021semantic} have proposed a Universal Transformer based semantic communication systems for text transmission with lower complexity.} Peng~\textit{et al.} \cite{peng2022robust} have designed robust semantic communication systems to prevent the semantic delivery from the source noise, e.g., typos and syntax errors. These existing works have revealed the potential of semantic communication in various intelligent tasks and moved towards the ideal intelligent communication. However, the proposed semantic communication systems only consider the current time slot inputs and ignore the previous time slots inputs. For the next step of task-oriented semantic communication, we can find some inspiration from human communication.

In the human communication \cite{tomasello2010origins}, people can perform the both memoryless tasks and memory tasks. Memoryless tasks are only relevant to the inputs received in the current time slot, e.g., receiving the image and recognizing its category. Memory tasks \cite{sukhbaatar2015end} are relevant to inputs received in both the current and past time slots, e.g., the response in the conversation relying not only on the currently listened sentences but also on the previous context.  \textcolor{black}{While the developed semantic communication systems only consider the inputs in the current time slot and neglected those in the previous time slots. With such a design, semantic communication is incapable of serving the memory tasks, such as scenario question answer, \textcolor{black}{scenario visual question task}, and scenario conversations.}   

\textcolor{black}{Note that one of the key modules in human communications is \textit{memory}, which can store the context semantic information and enable people to perform tasks requiring memory. We are inspired to introduce the memory module in semantic communications so as to execute tasks with and without memory.} By dosing so, machine to machine commendations and human to machine communications will become more intelligent and human-like, which could fully exploit the advancements of semantic communications. It is of great interest to design a semantic communication system that utilizes memory information to facilitate the semantic information transmission and task execution at the receiver. To design a semantic communication system with memory, we are facing the following challenges:
\begin{enumerate}
    \item[\textit{Q1}:] \textit{How to design the semantic-aware transceiver with memory module?}
    \item[\textit{Q2}:] \textit{How to ensure the effectiveness of transmitting memory over multiple time slots?}
\end{enumerate}

In this paper, we investigate a semantic communication for memory tasks by taking the scenario question answer task as an example. Particularly, we develop a DL enabled semantic communication system with memory (Mem-DeepSC) to address the aforementioned challenges. The main contributions of this paper are summarized as follows:
\begin{itemize}
    \item Based on the universal Transformer \cite{DehghaniGVUK19}, a transceiver with a memory module is proposed. In the proposed Mem-DeepSC, the transmitter can extract the semantic features at the sentence level effectively and the receiver can process received semantic features from the previous time-slots by employing the memory module, which addresses the aforementioned \textit{Q1}.
    \item To make the Mem-DeepSC applicable to \textcolor{black}{dynamic transmission environment}, \textcolor{black}{the relationship between the length of semantic signal and the channel noise is derived.}  Especially, two dynamic transmission methods are proposed to preserve semantic features from distortion and reduce the communication resources. Two lower bounds of mutual information are derived to train the dynamic transmission methods.  This addresses the aforementioned \textit{Q2}.
\end{itemize}

The rest of this paper is organized as follows. The system model is introduced in Section II.  The semantic communication system with memory module is proposed in Section III. Section IV details the proposed dynamic transmission methods. Numerical results are presented in Section V to show the performance of the proposed frameworks. Finally, Section VI concludes this paper.

\textit{Notation}:  Bold-font variables denote matrices or vectors. $\mathbb{C}^{n \times m}$ and $\mathbb{R}^{n \times m}$  represent complex and real matrices of size $n\times m$, respectively. $x \sim {\cal CN}(\mu,\sigma^2)$ means variable $x$ follows a circularly-symmetric complex Gaussian distribution with mean $\mu$ and covariance $\sigma^2$. $(\cdot)^{\text T}$ and $(\cdot)^{\text H}$ denote the transpose and Hermitian, respectively.  ${\bf I}_{M}$ is the $M\times M$ is the identity matrix, ${\bf I}_{M\times 1}$ is the all-one vector with length $M$, ${\bf 1}_{i}$ is the one-hot vector with one in the $i$-th position. $\odot$ and $\oslash$ are the element-wise multiplication and division, respectively. \textcolor{black}{${\bm x}(k)$ represents the signal at the $k$-th time slot. ${\bm x}[k]$ represents the $k$-th element in the vector. ${\bf X}[k]$ represents the $k$-th row in the matrix.}

\section{System Model}
\begin{figure*}[!t]
    \centering
    \includegraphics[width=150mm]{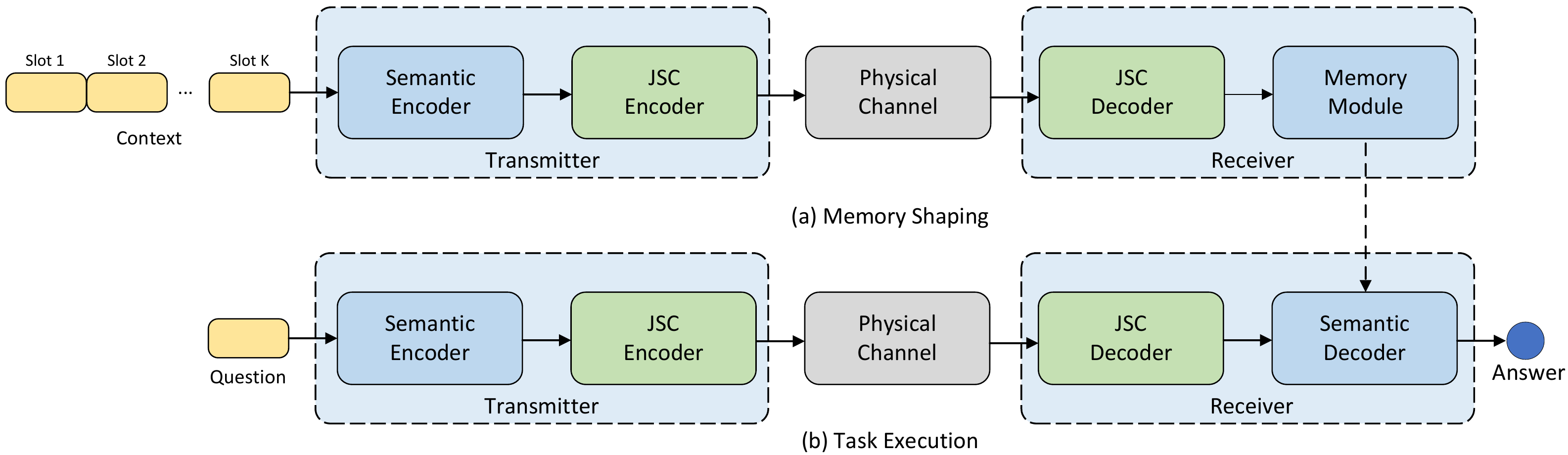}
    \caption{\textcolor{black}{The proposed framework for semantic communication systems with memory module.}}
    \label{fig:system-model}
\end{figure*}

As shown in Fig. \ref{fig:system-model}, we consider a single-input single-output (SISO) communication system, which is with one antenna at the transmitter and one at the receiver. We focus on the text scenario question answer, therefore, the transmission includes two phases: i) memory shaping to transmit the context, e.g., multiple sentences, images, or speeches, to the receiver via multiple time slots; ii) task execution to transmit the question relevant to the context so as to obtain the answer at the receiver. In this paper, we consider multiple sentences as the context.

The transceiver has three modules, a semantic codec to extract the semantic features of the source and perform the task, a joint source-channel (JSC) codec to compress and recover the semantic features, and the memory module to store the received context in multiple time slots and aid semantic decoder in performing the task.

\subsection{Memory Shaping}
We assume the $k$-th context, ${\bm s}(k)$, is transmitted during the $k$-th time slot and denote ${\bm s}^{c}$ and ${\bm s}^q$ as the context sentence and question sentence, respectively. In the memory shaping phase, the transmitter sends the context, e.g., multiple sentences, images, or speeches, to the receiver over multiple time slots.  Subsequently, with the semantic encoder and channel encoder, the $k$-th context sentence over the $k$-th time slot can be encoded as
\begin{equation}\label{eq1}
    {\bm x}^c(k) = C\left( S\left( {\bm s}^c(k) ; {\bm \alpha} \right); {\bm \beta} \right), 
\end{equation}
where ${\bm x}^c(k) \in \mathbb{C}^{L\times1}$ is the transmitted signals after the power normalization, $S\left(\cdot; {\bm \alpha} \right)$ and $C\left(\cdot; {\bm \beta}\right)$ are denoted as the semantic encoder with parameter ${\bm \alpha}$ and channel encoder with parameter ${\bm \beta}$, respectively. 

Transmitting the signals over the channels, the received signal can be presented as
\begin{equation}\label{eq2}
     {\bm y}^c{(k)} = {\bm h}(k)  \odot {\bm x}^c(k) + {\bm n}(k), 
\end{equation}
where $ {\bm h}(k) $ is the channel coefficients and $ {\bm n}$ is the additive white Gaussian noise (AWGN), in which ${\bm n}(k) \sim \mathcal{CN}\left(0, \sigma_n^2{\bf I}_L\right)$. For the Rayleigh fading channel, the channel coefficient follows ${\bm h}(k) \sim {\cal CN}\left(0,{\bf I}_{L}\right)$; for the Rician fading channel, it follows ${{\bm h}(k) \sim \cal CN}\left(\mu_h{\bf I}_{L\times 1},\sigma_h^2{\bf I}_{L}\right)$ with $\mu_h = \sqrt{r/(r+1)}$ and $\sigma_h = \sqrt{1/(r+1)}$,  where $r$ is the Rician coefficient. The SNR is defined as $\mathbb{E}({{{\left\| {\bm h}(k) \odot {\bm x}^c(k) \right\|}^2}}) /\mathbb{E}({{{\left\| \bm n(k) \right\|}^2}})$. 

With the estimated channel state information (CSI), $\hat{\bm h}$, the transmitted signals, $\hat{{\bm x}}(k)$, can be detected by
\begin{equation}\label{eq3}
    \hat {\bm x}^c(k) = {\hat {\bm h}}^{\tt H}(k) \odot {\bm y}^c(k) \oslash \left( {\hat {\bm h}}(k) \odot {\hat {\bm h}^{\tt H}}(k) \right).
\end{equation}
After signal detection, the semantic features can be recovered by
\begin{equation}\label{eq4}
    {\hat {\bm z}}^c(k) = C^{-1}\left( \hat {\bm x}^c(k); {\bm \gamma}\right),
\end{equation}
where $\hat {\bm z}^c{(k)}\in \mathbb{R}^{N \times1}$ and $C^{-1}\left(\cdot;{\bm \gamma} \right)$ is denoted as the channel decoder with parameter ${\bm \gamma}$. Then, the recovered semantic features will be inputted into the memory module.

\textcolor{black}{Inspired by the short-term memory, we model the memory module as a queue with length $T$, which only concerns the context. The memory queue at the $k$-th time slot is represented by
\begin{equation}\label{eq5}
    {\mathbf{Q}{(k)}} = [ \hat {\bm z}^c{(k-T + 1)}; \hat {\bm z}^c{(k-T +2)}; \cdots, \hat {\bm z}^c(k)].
\end{equation}
where ${\mathbf{Q}{(k)}} \in \mathbb{R}^{N \times T}$, $T$ is the length of memory. For $k<0$, $\hat {\bm z}^c{(k)}$ is the vector with all zero elements.  The memory queue is updated with the incoming received latest semantic features and pop the oldest features out of the queue.   For example, the memory queue at the $(k+1)$-th time slot is given by
\begin{equation}\label{eq1-3b}
    {\mathbf{Q}{(k+1)}} = [ \hat {\bm z}^c{(k-T)}; \hat {\bm z}^c{(k-T +1)}; \cdots, \hat {\bm z}^c{(k+1)}].
\end{equation}}

\textcolor{black}{For scenario communications, it is important to recognize the sequence of the memory queue. In other words, we need to know the order of sentences happened earlier or later. Therefore, we need to add the temporal information before inputting to the models, which is given by 
\begin{equation}\label{eq-mem-queue}
    {\mathbf{M}{(k)}} = {\mathbf{Q}{(k)}} + \mathbf{T},
\end{equation}
where $\mathbf{T}=[{\bm t}_1; {\bm t}_2; \cdots; {\bm t}_T]\in \mathbb{R}^{N \times T}$ is the temporal information matrix, any two elements of which satisfy $<{\bm t}_m, {\bm t}_n>=g(m-n)$ for some function $g(\cdot)$. We choose the positional coding employed in Transformer \cite{VaswaniSPUJGKP17} here.}

\textcolor{black}{When performing the memory task, the memory shaping phase will introduce the overheads in terms of time availability due to occupying multiple time slots. For example, considering the task offloading scenario, compared to the memoryless tasks that only use several time slots, the memory shaping will occupy multiple time slots and may block the transmission pipe. This makes the time consumption a little high at the initial memory shaping phase. However, when the memory module is shaped, such consumption can be alleviated by reusing or partly updating the shaped memory module, because the memory module serves multiple questions. Therefore, the frequency of carrying memory shaping phase depends on the demands of users. When performing the memory task, the user will decide whether the current context stored in the memory module can answer the question. If not, the memory needs to be updated.}

\subsection{Task Execution}
In the task execution phase, the transmitter sends the question sentence, ${\bm s}^q$, to the receiver to perform the task. Specially,  ${\bm s}^q$ is encoded into ${\bm x}^q$ by \eqref{eq1}, transmitted over the air, and decoded into $\hat {\bm z}^q$ by \eqref{eq4}. In the scenario question answer task, the question is not only relevant to only one context sentence but also multiple context sentences. Therefore, the answer is predicted with the question and memory together, which is represented as
\begin{equation}
    \textcolor{black}{{\hat a} = S^{-1}\left(\left[ \hat {\bm z}^q, \mathbf{M}{(k)}\right] ; {\bm{\varphi}}\right),}
\end{equation}
where $S^{-1}(\cdot;{\bm{\varphi}})$ is the semantic decoder with parameters ${\bm{\varphi}}$.

\section{Semantic Communication System with Memory}
In this section, we design a semantic communication system with memory, named Mem-DeepSC, to perform scenario question answer task, in which the universal Transformer is employed for text understanding.

\begin{figure*}[!t]
    \centering
    \includegraphics[width=180mm]{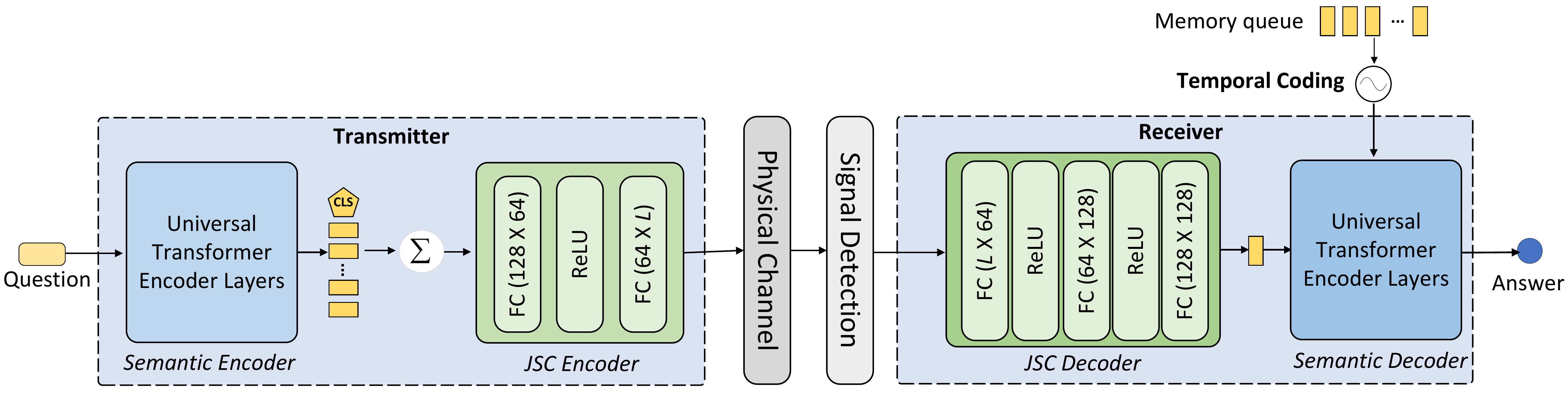}
    \caption{\textcolor{black}{The proposed network structure of Mem-DeepSC system.}}
    \label{fig:mem-deepsc}
\end{figure*}

\subsection{Model Description}
The proposed Mem-DeepSC is shown in Fig. \ref{fig:mem-deepsc}. The semantic encoder consists of universal Transformer encoder layer  with variable steps to extract the semantic feature of each word. In order to reduce the  transmission overheads, \textit{the summation operation} is taken here, in which these semantic features at the word level are merged to get one semantic feature at the sentence level.  The reason that we choose universal Transformer in the semantic codec can be summarized as follows:
\begin{itemize}
    \item \textcolor{black}{The state-of-arts, i.e., BERT \cite{DevlinCLT19} and GPT-3 \cite{brown2020language}, choose Transformer to deal with textual information, which shows the powerful extraction capability of text semantic information.}
    \item The universal Transformer can be trained and tested much faster than the architectures based on recurrent layers due to the parallel computation \cite{DehghaniGVUK19}.
    \item Compared with the classic Transformer, the universal Transformer shares the parameters, which can reduce the model size.
\end{itemize}

\textcolor{black}{After the semantic encoder, the JSC encoder employs multiple dense layers to compress the sentence semantic feature.} The reasons that we mainly use dense layer in the channel codec can be summarized as follows:
\begin{itemize}
    \item \textcolor{black}{The JSC codec aims to compress the semantic features and transmit it effectively over the air.  Compared with the CNN layer to capture the local information, the dense layer is good at capturing the global information and preserving the entire attributes, which follows the target of the JSC codec. This can enhance the system's robustness to channel noise.}
\end{itemize}

At the receiver, the JSC decoder correspondingly includes multiple dense layers to decompress sentence semantic feature and reduce the distortion from channels. The semantic decoder also contains the universal Transformer encoder layer with variable steps to find the relationship between the memory queue and the query feature to get the answer. Especially, the memory queue does not contain the temporal information inside. Therefore, \textit{temporal coding} is employed to add temporal information to the memory queue, in which we adopt the positional coding~\cite{VaswaniSPUJGKP17} as the temporal coding.

\begin{algorithm}[!t]
\caption{Mem-DeepSC Training Algorithm.}
\label{alg-1}
\SetKwInput{KwInput}{Input}                
\SetKwInput{KwInitia}{Initialization}
\SetKwInput{KwOutput}{Output}              
\SetKwInput{KwRet}{Return}
\DontPrintSemicolon

\SetKwFunction{FMain}{Main}
\SetKwFunction{FSE}{Train Semantic Codec}
\SetKwFunction{FCC}{Train  {JSC} Codec}
\SetKwFunction{FWN}{Train Whole Network}

  \SetKwProg{Fn}{Function}{:}{}
  \Fn{\FSE{}}{
        \KwInput{ $\left \{ ({\bm s}^c{(1)}, {\bm s}^c{(2)}, \cdots, {\bm s}^c{(K)}), {\bm s}^q, a \right\} $ from dataset.}
        \For{$k = 1 \to K$}
        {
   		${S\left( {{\bm s}^c{(k)};{\bm \alpha} } \right)}  \to  {\bf W}^c{(k)}$,\;
   		Take the summation operation, ${\bm z}^c(k) = \sum_j {\bf W}^c{(k)[j]}$,\;
   		}
   		$ {S\left( {{\bm s}^q;{\bm \alpha} } \right)}  \to  {\bf W}^q$, and ${\bm z}^q = \sum_j {\bf W}^q{[j]}$, \; 
   		Shape the memory queue ${\mathbf{Q}{(K)}}$ by \eqref{eq5},\;
   		Take the temporal coding for ${\mathbf{M}{(K)}}$ by \eqref{eq-mem-queue}, \;
   		$  S^{-1}\left(\left[ {\bm z}^q, \mathbf{M}{(K)}\right] ; {\bm{\varphi}}\right) \to  \hat a $, \; 
   		Compute CE loss with  $a$ and ${\hat a}$.\;
        Train ${\bm{\alpha }}, {\bm{\varphi}}$ $\to$ Gradient descent with CE loss.\;
        \KwRet{{${S\left( \cdot; {\bm \alpha }\right)}$ and ${S^{-1}\left( {\cdot;{{\bm{\varphi }}}}\right)}$.}} 
  }

  \SetKwProg{Fn}{Function}{:}{}
  \Fn{\FCC{}}{
        \KwInput{Semantic features ${\bm z}^c(k)$.}
        \textbf{Transmitter}:\;
   		\quad $ {C\left( {{\bm z}^c{(k)};{{\bm{\beta }}}} \right)}  \to  {\bm x}^c{(k)} $, \; 
   		\quad {Power Normalization,} \;
   		\quad Transmit ${\bm x}^c{(k)} $  over the air.\;
   		\textbf{Receiver}:\;
   		\quad Receive ${\bm y}^c(k)$, \;
   		\quad Signal detection by \eqref{eq3} to get ${\hat {\bm x}}^c{(k)}$,\;
   		\quad $ {C^{-1}\left( {\hat {\bm x}^c{(k)};{\bm{\gamma }}} \right)} \to  {\hat {\bm z}}^c{(k)} $, \; 
   		Compute MSE loss with ${{\bm z}}^c{(k)}$ and ${\hat {\bm z}}^c{(k)}$.\;
        Train ${{\bm{\beta }}}, {{\bm{\gamma }}}$ $\to$ Gradient descent with MSE loss.\;
        \KwRet{${C\left( {\cdot;{{\bm{\beta }}}} \right)}$ and ${C^{-1}\left( {\cdot;{{\bm{\gamma }}}} \right)}$.}
  }
 
  \SetKwProg{Fn}{Function}{:}{}
  \Fn{\FWN{}}{
         \KwInput{ $\left \{ ({\bm s}^c{(1)}, {\bm s}^c{(2)}, \cdots, {\bm s}^c{(K)}), {\bm s}^q, a \right\} $ from dataset.}
        Repeat line 2-5, 12-19, and 6-8 to get $\hat a$,\;
  		Compute CE loss  with $\hat a$ and $a$. \;
        Train ${{\bm{\alpha }}}, {{\bm{\beta }}}, {{\bm{\gamma }}}, {{\bm{\varphi }}}$ $\to$ Gradient descent with CE loss.\;
        \KwRet{{${S\left( {\cdot;{{\bm{\alpha }}}}\right)}$, ${S^{-1}\left( {\cdot;{{\bm{\varphi }}}}\right)}$, ${C\left( {\cdot;{{\bm{\beta }}}} \right)}$, and ${C^{-1}\left( {\cdot;{{\bm{\gamma }}}} \right)}$.} }
  }
\end{algorithm}

\subsection{Training Details}
As shown in Algorithm \ref{alg-1}, the training of Mem-DeepSC includes three steps, which is similar to the training algorithm proposed in \cite{xie2021task}. The first step is to train the semantic codec. In order to improve the accuracy of answers, we choose the cross-entropy (CE) as the loss function instead of the answer accuracy. The cross-entropy is given by
\begin{equation}\label{eq7}
    {\cal L}_{\tt CE} = -p\left( a \right)\text{log}\left(p\left( \hat a \right)\right),
\end{equation}
where $p\left(a\right)$ is the real probability of answer and $p\left(\hat a\right)$ is the predicted probability. After convergence, the model learns to extract the semantic features and predict the answers. \textcolor{black}{The following proposition proven in Appendix A reveals the relationship between cross-entropy and the answer accuracy.}
\begin{prop}\label{prop1}
Cross entropy loss function is the refined function of answer accuracy and is more stable during training.
\end{prop}

\textcolor{black}{After converged, the model is capable of extracting semantic features and predicting the answer accurately.} Subsequently, the second step is to ensure the semantic features transmitted over the air effectively. Thus, the JSC codec is trained to learn the compression and decompression of the semantic features as well as to deal with the channel distortion with the mean-squared error (MSE) loss function,
\begin{equation}\label{eqmse}
    {\cal L}_{\tt MSE} = \left \| {\bm z}^c(k) - \hat {\bm z}^c(k) \right \|^2,
\end{equation}
where ${\bm z}^c(k)$ and $\hat {\bm z}^c(k)$ are the original semantic features and the recovered semantic features, respectively. 

Finally, the third step is to optimize the entire system jointly to achieve the global optimization. The semantic codec and JSC codec are trained jointly with the CE loss function to reduce the error propagation between each module.  

With the Mem-DeepSC, the memory-related tasks can be performed. However, the context is transmitted via multiple time slots. If each time slot has different channel conditions, the damage to the semantic information is inevitable at the worse channel conditions, which affects the prediction accuracy. Therefore,  in order to preserve the semantic information and save the communication overheads over multiple time slots, we further develop an adaptive rate transmission method.

\section{Adaptive Rate Transmission}
In this section,  we firstly derive \textcolor{black}{the relationship between the length of semantic signal and channel noise,} which inspires us to transmit different length signals according to SNRs. We then develop two dynamic transmission methods, importance mask and consecutive mask for saving the communication resources and preventing the outage for memory transmission to different SNRs.

\subsection{The Relationship Between the Length of Semantic Signal and Channel Noise}

Adaptive modulation has been developed for conventional communications \cite{goldsmith2005wireless}, where the modulation order and code rate change according to SNRs. The same spirit can be used in semantic communications if there exists the relationship between the length of semantic signal and channel noise over AWGN. In this situation, we can achieve such adaptive operation by masking some elements, i.e., masking less at low SNR regimes to ensure the reliability of performing tasks and masking more elements at high SNR regimes to achieve a higher transmission rate.

How many semantic elements should be transmitted? The existing works \cite{Tung21, ShaoMZ22} employ neural networks to learn how to determine the number of transmitted semantic elements dynamically, which lacks of interpretability. Therefore, we provide a theoretical analysis of the relationship between the length of semantic signal and the channel noise to guide us to determine the number of semantic elements at certain SNR. 

The key is to find the relationship between the noise level and the number of elements that can be transmitted correctly. Firstly, we model ${\bm x}^c(k)$ into
\begin{equation}\label{eq13}
    {\bm x}^c(k) = {\bm r}^c(k) + {\bm n}_{\tt model},
\end{equation}
where ${\bm r}^c(k)$ is the semantic information selected from the latent semantic codewords, 
${\bm n}_{\tt model} \sim {\cal CN}\left(0, \sigma^2_m {\bf I} \right)$ is the model noise. We generally initialize the model weights with Gaussian distribution and apply the batch normalization/layer normalization to normalize the outputs following ${\cal N}(0,1)$ \cite{ioffe2015batch}. Therefore, we model the model noise with Gaussian distribution. In deep learning, the model noise is caused by the unstable gradients descending, the training data noise, and so on. The model noise can be alleviated by the larger dataset, the refined optimizer, and the re-designed loss function but cannot be removed. 

\begin{figure*}[!t]
    \centering
    \includegraphics[width=160mm]{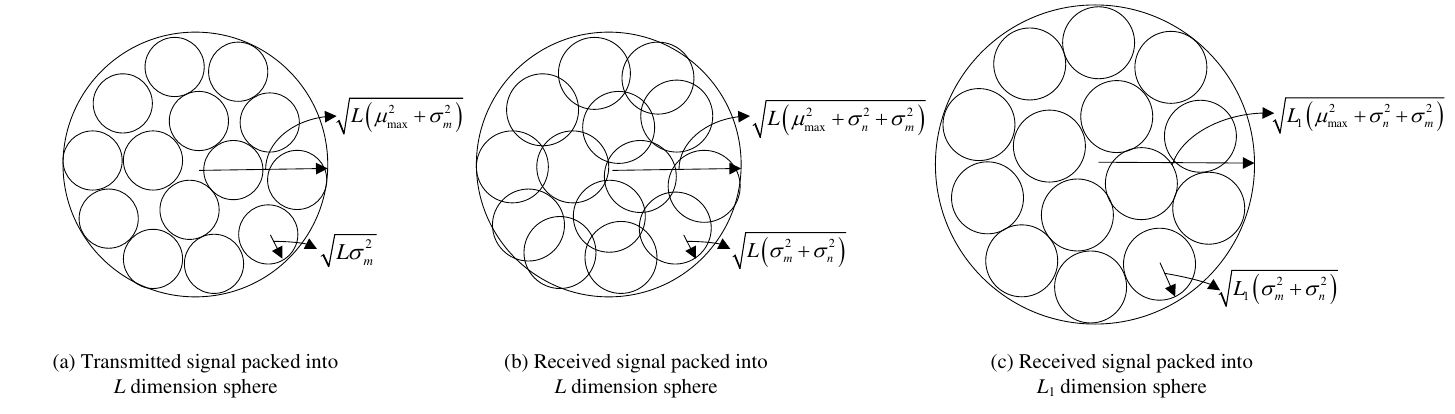}
    \caption{\textcolor{black}{An example of the relationship between the length of semantic signal and channel noise over AWGN: (a) the transmitted signal of length $L$ packed into $L$ dimension sphere; (b) the received signal of length $L$ over AWGN channels packed into $L$ dimension sphere; (c) the received signal of length $L_1$ over AWGN channels packed into $L_1$ dimension sphere.} }
    \label{fig:sphere-packing}
\end{figure*}

Assume the length of ${\bm x}^c(k)$ is $L$. By applying the packing sphere theory \cite{tse2005fundamentals}, {${\bm x}^c(k)$} can be mapped to the $L$-dimension sphere space as shown in Fig. \ref{fig:sphere-packing}(a). In the Fig. \ref{fig:sphere-packing}(a), the smaller sphere represents the  noise sphere with radius ${\sqrt{L}\sigma_m}$ and the larger sphere is the signal sphere with radius ${\sqrt{L(\mu^2_{\max} + \sigma_m^2)}}$, where $\mu_{\max}$ is the maximum value in the latent semantic codewords. The reason that noise spheres spread the signal sphere is that the latent semantic codewords have different constellation points. Communication is reliable as long as the noise spheres do not overlap. Therefore, there exists a minimum length of $L$ to prevent the overlap from the model noise. In other words, the number of semantic codewords that can be packed with non-overlapping noise sphere over the model noise is
\begin{equation}\label{eq14}
    N = \frac{\left( \sqrt{L\left(\mu^2_{\max} + \sigma_m^2\right)} \right)^L}{\left( \sqrt{L\left( \sigma_m^2\right)} \right)^L}= {\left( {1 + \frac{{{\mu^2_{\max }}}}{{\sigma _m^2}}} \right)^{\frac{L}{2}}}.
\end{equation}

After transmitting ${\bm x}^c(k)$ over the AWGN channels, the received signals can be represented by submitting \eqref{eq13} into \eqref{eq2},
\begin{equation}\label{eq15}
    {\bm y}^c(k) = {\bm r}^c(k) + {\bm n}_{\tt model} + {\bm n}_{\text{channel}},
\end{equation}
where ${\bm n}$ in \eqref{eq2} is re-denoted to ${\bm n}_{\text{channel}}$. The ${\bm y}^c(k)$ can also be mapped to the $L$-dimension sphere space shown in Fig.~\ref{fig:sphere-packing}(b). Because of the channel noise, the radius of noise sphere increases from ${\sqrt{L(\sigma_m^2)}}$ to ${\sqrt{L(\sigma_n^2 + \sigma_m^2)}}$, which makes the noise spheres overlap. 

In order to eliminate the overlapping, one way is to increase the length of ${\bm x}^c(k)$ from $L$ to $L_1$ ($L_1 > L$) to enlarge the volume of the signal sphere so that the enlarged noise spheres do not overlap. Then, the number of semantic codewords that can be packed with non-overlapping noise sphere over the model noise and the channel noise is  
\begin{equation}\label{eq16}
    N = \frac{\left( \sqrt{L_1\left(\mu^2_{\max} + \sigma_m^2 + \sigma_n^2\right)} \right)^{L_1}}{\left( \sqrt{L_1\left( \sigma_m^2 + \sigma_n^2\right)} \right)^{L_1}}= {\left( {1 + \frac{{{\mu^2_{\max }}}}{{\sigma_m^2 + \sigma_n^2}}} \right)^{\frac{L_1}{2}}}.
\end{equation}

The semantic codewords only describe the semantic information of the source and are irrelevant to the channel noise, which means that the numbers of semantic codewords in \eqref{eq14} and \eqref{eq16} are the same. Therefore, the relationship between $L$ and $L_1$ can be derived as shown in proposition \ref{prop-4}.

\begin{prop}\label{prop-4}
\textcolor{black}{Given the minimum length, $L$, to prevent from model noise, the minimum length for reliable communication over AWGN channels is 
\begin{equation}\label{eq17}
{L_1} = L \times \frac{{\log \left( {1 + \frac{{\mu _{\max }^2}}{{\sigma _m^2}}} \right)}}{{\log \left( {1 + \frac{{\mu _{\max }^2}}{{\sigma _m^2 + \sigma _n^2}}} \right)}}.
\end{equation}}
\end{prop}

With proposition \ref{prop-4}, the masked ratio to different SNRs can be computed theoretically. With \eqref{eq17}, the asymptotic analysis can be derived into four cases listed below.

\begin{itemize}
    \item \textbf{Case 1}: When $\sigma^2_n \to 0$, \textcolor{black}{then $L_1 \to L$.} The number of transmitted symbols will converge to minimum $L$. In this case, the semantic communication system can be viewed as the compressor and decompressor.
    \item \textbf{Case 2}: When $\sigma^2_n \to \infty$, \textcolor{black}{then $L_1 \to \infty$.} The number of transmitted symbols will lead to infinity. In this case, the semantic communication system experiences an outage. 
    \item \textbf{Case 3}: When $\sigma^2_m \to 0$, then $L \to 0$. \textcolor{black}{$L_1$} only depends on the channel noise and can be computed by
    \begin{equation}
        \textcolor{black}{L_1 = \frac{2\log \left( N\right)}{{\log \left( {1 + \frac{{\mu _{\max }^2}}{{ \sigma _n^2}}} \right)}}.}
    \end{equation}
    In this situation, \textcolor{black}{$L_1$} is computed by the traditional channel capacity and the number of semantic codewords. 
    \item \textbf{Case 4}: When $\sigma^2_m \to \infty$, then $L \to \infty$. The semantic communication system experiences an outage, similar to case 2.
\end{itemize}

\begin{figure*}[!t]
    \centering
    \includegraphics[width=180mm]{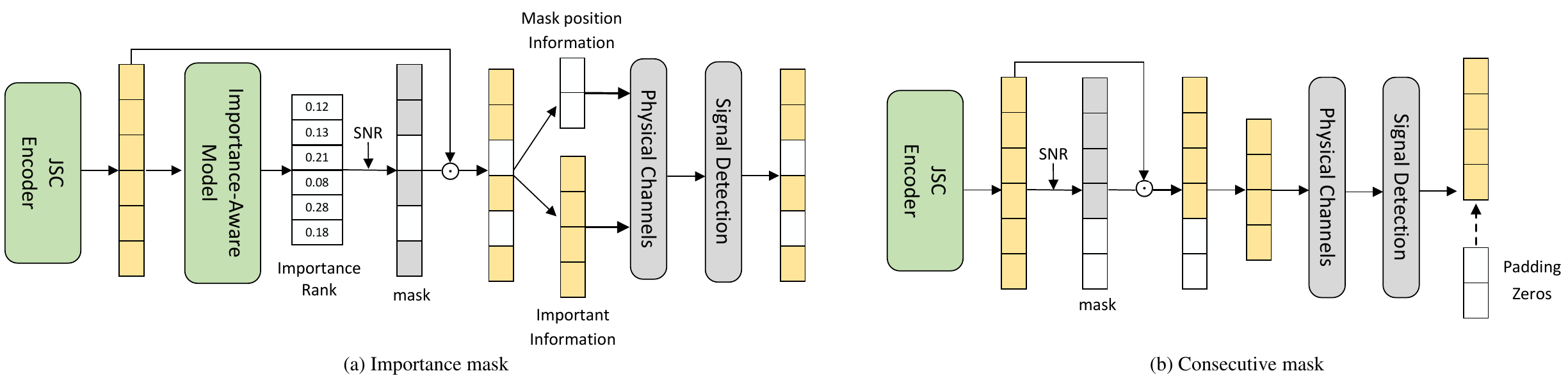}
    \caption{\textcolor{black}{The proposed two dynamic transmission methods: (a) Importance mask method; (b) Consecutive mask method.}}
    \label{fig:mask-methods}
\end{figure*}
The key differences between the relationship and traditional channel capacity can be summarized in the following,
\begin{itemize}
    \item The relationship indicates how much semantic information can be transmitted error-free while the traditional channel capacity indicates how many bits can be transmitted error-free.
    \item The length of semantic vectors is affected by three points, 1) the number of semantic codewords, 2) the model noise, and 3) the channel noise. But the channel capacity only depends on the channel noise. 
    \item When channel noise disappears, the length of semantic vectors has the lower bound, $L$. The traditional channel capacity does not have such a lower bound.
\end{itemize}

With the relationship, it is possible to achieve dynamic transmission. The key to achieving such a dynamic transmission in semantic communication systems is to identify which elements are more important than the others and mask the unimportant ones. As shown in Fig. \ref{fig:mask-methods}, we propose two mask methods subsequently, importance mask method and consecutive mask method.

\subsection{Importance Mask} 
As shown in Fig. \ref{fig:mask-methods}(a), the importance mask method introduces the importance-aware model to identify the importance order among the elements of ${\bm x}^c(k)$, which can be expressed as
\begin{equation}
    {\bm p}^c(k) = F\left({\bm x}^c(k); {\bm \theta}\right),
\end{equation}
where $F\left(\cdot; {\bm \theta}\right)$ is the importance-aware model with learnable parameter ${\bm \theta}$, ${\bm p}^c(k)$ is the importance rank of ${\bm x}^c(k)$, in which the bigger value means that the corresponding element is more important.

By setting the threshold, $\kappa$, the mask, ${\bm m}^c(k)$, can be computed with the ${\bm p}^c(k)$ by
\begin{equation}\label{eq10}
    {\bm m}^c{(k)[i]} = \left\{ 
        \begin{array}{l}
            1,\, {\bm p}^c{(k)[i]} > \kappa, \\
            0,\, {\bm p}^c{(k)[i]} \le \kappa.
        \end{array} \right.
\end{equation}
\textcolor{black}{where $\kappa$ can be determined by the $L_1$-th element in the sorted importance rank with the descend order, and $L_1$ is computed by \eqref{eq17}.}

Then, the masked transmitted signal can be generated by
\begin{equation}\label{eq11}
    \tilde {\bm x}^c(k) = {\bm x}^c(k) \odot {\bm m}^c(k).
\end{equation}

With $\tilde {\bm x}^c(k)$,  the transmitter can send the only non-zero elements and the position information of zero elements to reduce the communication overheads.  

After transmitting $\tilde {\bm x}^c(k)$ over the air, the receiver follows the same processing to perform signal detection, JSC decoding, and semantic decoding.

\subsubsection{Loss Function Design}
In order to train the importance model, the optimization goal is to keep more information related the task in the masked signals to prevent performance degradation. Therefore, the mutual information between $\tilde {\bm x}^c(k)$ and the goal $a$ is employed as the loss function,
\begin{equation}\label{eq-12}
    {\cal L}_{\tt MI} = -I\left(\tilde{\bm x}^c(k); a\right).
\end{equation}

However, minimizing \eqref{eq-12} with gradients descending algorithm is hard since ${\cal L}_{\tt MI}$ is undifferentiable and difficult to compute. There are several methods to alleviate the problem, e.g., employing the mutual information estimator and the numerical approximation. Even if these methods solve the undifferentiable problem, it is still unstable in estimating the mutual information. In order to achieve stable optimization, an approximate bound-optimization (or Majorize-Minimize) algorithm is employed. The bound-optimization aims to construct the desired majorized/minorized version of the objective function. Following the idea, two propositions are proposed for the bound-optimization of mutual information, which are \textcolor{black}{proven} in Appendices B and C, respectively.

\begin{prop}\label{prop-2}
For classification tasks, alternately maximizing the mutual information can be viewed as a bound optimization of the cross entropy.
\end{prop}
\begin{prop}\label{prop-3}
For regression tasks, alternately maximizing the mutual information can be viewed as a bound optimization of the mean absolute error.
\end{prop}

With Propositions \ref{prop-2} and \ref{prop-3}, the mutual information loss function in \eqref{eq-12} can be changed to the cross-entropy loss function in \eqref{eq7}. 

\subsubsection{Training Details}
As shown in Algorithm \ref{alg-2}, the importance model is trained by the CE loss function and the frozen Mem-DeepSC model. The training importance model takes the backpropagations from the semantic decoder to guide the importance model, in which the SoftKMax activation function is employed to bridge the backpropagation from mask to importance model. In other words, the importance model can learn which elements have more contributions/importance to the task performance by minimizing the CE loss function.

\begin{algorithm}[!t]
\caption{Importance Mask Training Algorithm.}
\label{alg-2}
\SetKwInput{KwInput}{Input}                
\SetKwInput{KwInitia}{Transmitter}
\SetKwInput{KwOutput}{Output}              
\SetKwInput{KwRet}{Return}
\DontPrintSemicolon

\SetKwFunction{FMain}{Main}
\SetKwFunction{FSE}{Train Importance Model}

  \SetKwProg{Fn}{Function}{:}{}
  \Fn{\FSE{}}{
        \KwInput{ $\left \{ {\bm x}^c{(k)}, {\bm x}^q, a \right\}$. Freeze Mem-DeepSC.}
        
        \textbf{Transmitter}:\;
   		\quad ${F\left( {{\bm x}^c{(k)};{\bm \theta} } \right)}  \to  {\bm p}^c{(k)}$,\;
   		\quad Compute the mask, ${\bm m}^c(k)$, by \eqref{eq10}\;
   		\quad Compute the mask signal, $\tilde{\bm x}^c(k)$, by \eqref{eq11},\;
   		\quad Transmit $\tilde{\bm x}^c(k)$ and ${\bm x}^q$ over the air,\;
   		
   		\textbf{Receiver}:\;
   		\quad Receive signal and perform signal detection, \; 
   		\quad $ {C^{-1}\left( {\tilde {\bm x}^c{(k)};{\bm{\gamma }}} \right)} \to  {\hat {\bm z}}^c{(k)} $, and $ {C^{-1}\left( {\hat {\bm x}^q;{\bm{\gamma }}} \right)} \to  {\hat {\bm z}}^q $\;
   		\quad Update the memory queue, ${\mathbf{Q}{(k)}}$, with  $\hat {\bm z}^c(k)$,\;
   		\quad Take the temporal coding for ${\mathbf{M}{(k)}}$ by \eqref{eq-mem-queue}, \;
   		\quad $S^{-1}\left(\left[ \hat {\bm z}^q, \mathbf{M}{(k)}\right] ; {\bm{\varphi}}\right) \to  \hat a $, \; 
   		Compute CE loss with  $a$ and ${\hat a}$.\;
        Train ${\bm \theta}$ $\to$ Gradient descent with CE loss.\;
        \KwRet{{${F\left( \cdot; {\bm \theta }\right)}$.}} 
  }

\end{algorithm}

\subsection{Consecutive Mask}
As shown in Fig. \ref{fig:mask-methods}(b), the consecutive mask method masks the last consecutive elements in the ${\bm x}^c(k)$ to zero, so that the transmitter only sends the non-zero elements and the receiver pads the received signals with zeros to the same length of ${\bm x}^c(k)$. The consecutive mask method does not need to transmit the additional mask position information but to re-train the Mem-DeepSC model. Since the importance rank of the elements of ${\bm x}^c(k)$ is not consecutive, directly masking these consecutive elements may experience performance degradation. The Mem-DeepSC needs to be re-trained with the consecutive mask so that it can learn to re-organize the elements of ${\bm x}^c(k)$ following the order of descend importance. 

The training of the consecutive mask method only includes one step, which is similar to the \texttt{Train Whole Network} in Algorithm~\ref{alg-1} but with two additional operations, i.e., masking operation before transmitting and padding operation after signal detection. The loss function during the training is the CE loss function.

\section{Simulation Results}
In this section, we compare the proposed semantic communication systems with memory with the traditional source coding and channel coding method over various channels, in which the proposed mask methods are compared with different benchmarks.

\subsection{Implementation Details}
\subsubsection{The Dataset} 
We choose the \textit{bAbI-10k} dataset \cite{WestonBCM15}, including 20 different types of scenario tasks. Each example is composed of a set of facts, a question, the answer, and the supporting facts that lead to the answer. We split the 10k examples into 8k examples for training, 1k examples for validation, and 1k examples for testing.

\subsubsection{Traing Settings}
The semantic encoder and decoder consist of the universal Transformer encoder layer with 3 steps and with 6 steps, respectively, in which the width of the layer is 128. The importance model is composed of one Transformer encoder layer with the width of 64. The other training settings are listed in Table \ref{tab:1}.

\subsubsection{Benchmarks and Performance Metrics}
We adopt the typical source and channel coding method as the benchmark of the proposed Mem-DeepSC, and the random mask method as the counterpart of the proposed two mask methods. 
\begin{itemize}
    \item \textcolor{black}{Separate Mem-DeepSC: The semantic codec and channel codec are trained with \eqref{eq7} and \eqref{eqmse}, respectively.}
    \item Conventional methods: To perform the source and channel coding separately, we use the following technologies, respectively:
    \begin{itemize}
        \item 8-bit unicode transformation format (UTF-8) encoding for text source coding, a commonly used method in text compression;
        \item Turbo coding for text channel coding, popular channel coding for a small size file;
        \item 16-quadrature amplitude modulation as the modulation.
    \end{itemize}
    \item Random Mask: Mask the elements in the transmitted signal randomly.
\end{itemize} 
In the simulation, the coding rate is 1/3 and the block length is 256 for the Turbo codes. The coherent time is set as the transmission time for each context in the simulation. We set $r=2$ for the Rician channels and ${\bf h} = \bf 1$ for the AWGN channels. In order to compute the required length of semantic vectors, we train multiple Mem-DeepSC with different sizes to find the values of $\mu_{\max}$ and $\sigma^2_m$. For Mem-DeepSC, $\mu_{\max} =1$ and  $\sigma^2_m = 1.44$. Answer accuracy is used as the metric to compute the ratio between the number of correct answers and that of all generated answers.

\begin{table}[!t]
\caption{The Training Settings.}
\label{tab:1}
\footnotesize
\centering
\begin{tabular}{ c|c|c|c } 
\toprule
& Batch Size & Learning Rate & Epoch \\
\midrule
\texttt{Train Semantic Codec}& 200 &  $5\times 10^{-4}$ & 250\\
\midrule
\texttt{Train Channel Codec} & 100  & $1\times 10^{-4}$ & 50 \\
\midrule
\texttt{Train Whole Network} & 200 & $5\times 10^{-4}$ & 30  \\
\midrule
\texttt{Train Importance Mask} & 200 & $5\times 10^{-4}$ & 10  \\
\midrule
\texttt{Train Consecutive Mask} & 200 & $1\times 10^{-4}$ & 50 \\
\bottomrule
\end{tabular}
\end{table}

\begin{figure}[!t]
	\centering
	\subfigure[AWGN channels.]{
			\includegraphics[width=75mm]{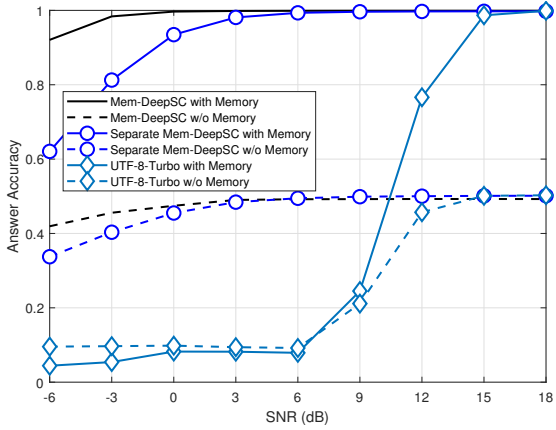}
		\label{fig:mem-deepsc-1}
	}
	
    	\subfigure[Rician channels with perfect CSI.]{
		 	\includegraphics[width=75mm]{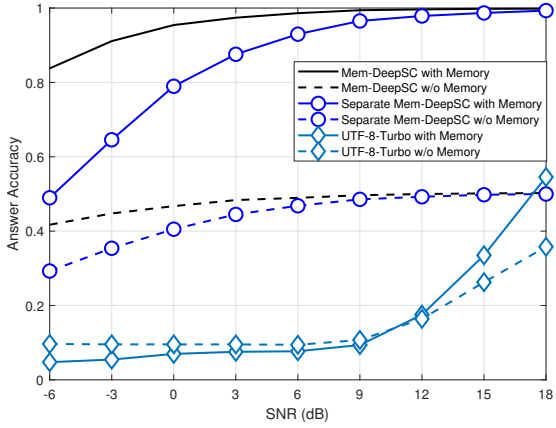}
		\label{fig:mem-deepsc-2}
    }
    
    \subfigure[Rician channels with imperfect CSI.]{
   		 	\includegraphics[width=75mm]{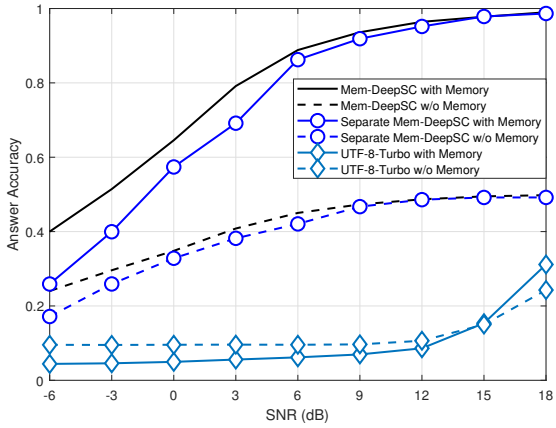}
		\label{fig:mem-deepsc-3}
    }
	\caption{Answer accuracy comparison between Mem-DeepSC and UTF-8-Turbo with 16-QAM over different channels.}
	\label{fig:mem-deepsc-simulation}
\end{figure}

\subsection{Memory Semantic Communication Systems}
Fig. \ref{fig:mem-deepsc-simulation} compares the answer accuracies over different channels, in which the Mem-DeepSC and the UTF-8-Turbo transmit 32 symbols per sentence and 190 symbols per sentence, respectively. The proposed Mem-DeepSC with memory outperforms all the benchmarks at the answer accuracy in all SNR regimes by the margin of 0.8. Compared the Mem-DeepSC with memory and without memory, the memory module can significantly improve the answer accuracy, which validates the effectiveness of the memory module in memory-related transmission tasks. \textcolor{black}{Besides, the Mem-DeepSC outperforms the separate Mem-DeepSC in low SNR regimes, which means that the three stage training algorithm can help improve the robustness to channel noise.} From the AWGN channels to the Rician channels, the proposed Mem-DeepSC with memory experiences slight answer accuracy degradation in the low SNR regimes but the UTF-8-Turbo has an obvious performance loss in all SNR regimes. The inaccurate CSI deteriorates the answer accuracy for both methods, however, the proposed Mem-DeepSC can keep a similar answer accuracy in high SNR regimes, which shows the robustness of the proposed Mem-DeepSC.

\begin{figure}[!t]
	\centering
	\subfigure[AWGN channels.]{
			\includegraphics[width=75mm]{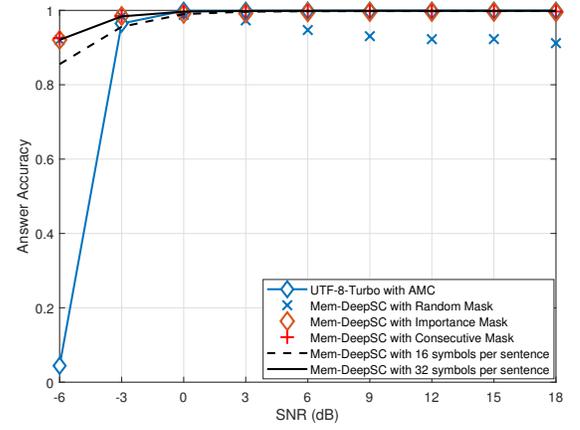}
		\label{fig:dynamic-1}
	}
	
    	\subfigure[Rician channels with perfect CSI.]{
		 	\includegraphics[width=75mm]{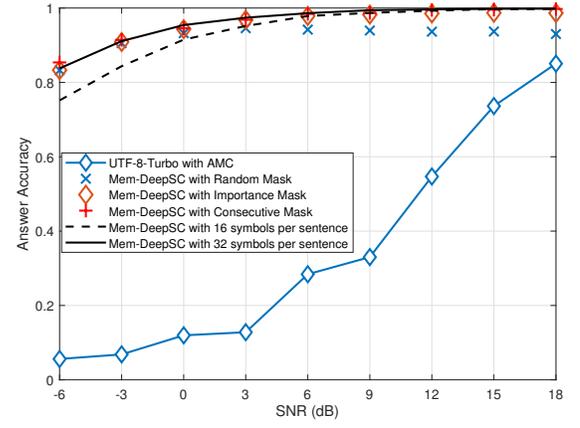}
		\label{fig:dynamic-2}
    }
    
    \subfigure[Rician channels with imperfect CSI.]{
   		 	\includegraphics[width=75mm]{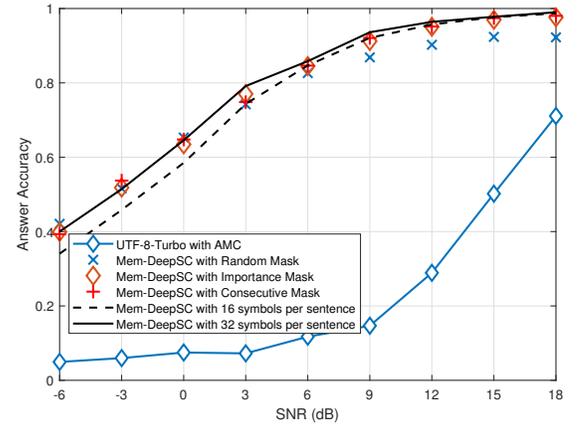}
		\label{fig:dynamic-3}
    }
	\caption{\textcolor{black}{Answer accuracy comparison between Mem-DeepSC with different number of transmitted symbols over different channels.}}
	\label{fig:dynamic}
\end{figure}

\subsection{The Proposed Mask Methods}
Table \ref{tab:2} compares the number of transmitted symbols for different methods.  \textcolor{black}{Compared to the UTF-8-Turbo with the adaptive modulation and channel coding (AMC), the proposed Mem-DeepSC decreases the number of the transmitted symbols significantly with only 4\%-16.8\% symbols.} The reason is that the Mem-DeepSC transmits the semantic information at the sentence level instead of at the letter/word level.  Besides, applying the dynamic methods can further reduce the number of transmitted symbols from 32 symbols to 16 symbols per sentence as the SNR increases, especially saving an additional 50\% symbols in the high SNR regimes. Then, the effectiveness of \eqref{eq17} is validated by the following simulation in Fig. \ref{fig:dynamic}.

Fig. \ref{fig:dynamic} verifies the effectiveness of the proposed mask strategy. \textcolor{black}{For Mem-DeepSC with no mask, we provided two cases with 16 symbols and 32 symbols per sentence, respectively. Utilizing adaptive modulation and channel coding (AMC) on UTF-8-Turbo can yield comparable answer accuracy to that of Mem-DeepSC over AWGN channels. However, this comes at the expense of a reduced transmission rate. }Then comprising the no mask cases with different number of symbols per sentence, increasing the number of symbols per sentence leads to higher answer accuracy in low SNR regimes but the gain disappears as the SNR increases. This suggested that the semantic communication systems can employ more symbols in low SNR to improve the robustness and transmit fewer symbols in the high SNR regimes to improve the transmission efficiency. The proposed importance mask and consecutive mask keep the similar answer accuracy as the Mem-DeepSC with 32 symbols per sentence in all SNR regimes over the AWGN and the Rician channels.

\begin{table*}[!t]
\caption{The number of transmitted symbols comparison between different methods.}
\label{tab:2}
\centering
\begin{tabular}{ c|c|c|c|c|c } 
\toprule
& \multicolumn{5}{c}{Number of Transmitted Symbols} \\
\midrule
& -6dB  & 0dB & 6dB  & 12dB & 18dB\\
\midrule
Mem-DeepSC & \multicolumn{5}{c}{32} \\
\midrule
Dynamic Transmission & 32 & 25 & 18 & 16 & 16  \\
\midrule
\makecell[c]{UTF-8-Turbo} & \multicolumn{5}{c}{190}   \\
\midrule
\makecell[c]{UTF-8-Turbo with AMC \\(AWGN Channels)} & \makecell[c]{760 \\(BPSK)} & \makecell[c]{760 \\(BPSK)} & \makecell[c]{380 \\(4QAM)} & \makecell[c]{253 \\(8QAM)}  & \makecell[c]{190\\(16QAM)}   \\
\midrule
\makecell[c]{UTF-8-Turbo with AMC \\(Rician Fading Channels)} & \makecell[c]{760  \\(BPSK)} & \makecell[c]{760  \\(BPSK)} & \makecell[c]{380  \\(4QAM)}  & \makecell[c]{253 \\(8QAM)} &  \makecell[c]{253 \\(8QAM)} \\
\bottomrule
\end{tabular}
\end{table*}

However, the random mask experiences significant answer accuracy loss in the high SNR regimes as the number of masked elements increases because some important elements are masked randomly to reduce the answer accuracy. This validates the effectiveness of both proposed mask methods.

\section{Conclusion}
In this paper, we have proposed a memory-aided semantic communication system, named Mem-DeepSC. The scenario question answer task is taken as the example. The Mem-DeepSC can extract the semantic information at the sentence level to reduce the number of the transmitted symbols and deal with the context information at the receiver by introducing the memory queue. Moreover, with the memory module, the Mem-DeepSC can deal with the memory-related tasks compared to that without the memory module, which is closer to human-like communication. Besides, the relationship between the length of semantic signal and the channel noise over AWGN is derived to decide how many symbols are required to be transmitted at different SNRs. Two dynamic transmission methods are proposed to mask the unimportant elements in the transmitted signals, which can employ more symbols in the low SNR to improve the robustness and transmit fewer symbols in the high SNR regimes to improve the transmission efficiency.  In particular, the dynamic transmission methods can save an additional 50\%  transmitted symbols.  Therefore, the semantic communication system with memory is an attractive alternative to intelligent communication systems.

\appendices
\section{Proof of Proposition 1}
Given the mini-batch, $B$, the question-answer accuracy can be computed by
\begin{equation}\label{eq18}
    Acc=\frac{1}{{\left| B \right|}}\sum\limits_{ B} {\left\langle {{{\mathbf{1}}_i},{{\mathbf{1}}_j}} \right\rangle },
\end{equation}
where $|B|$ is the batch size, and ${\bm 1}_i$ is the one-hot vector with one in the $i$-th position, ${\bm 1}_i$ is the real answer with label $i$, and ${\bm 1}_j$ represents the predicted answer with predicted label $j$, which is computed by
\begin{equation}
    {\bf 1}_j = \text{onehot}(\arg \max({\bm l}))),
\end{equation}
where ${\bm l}$ is the output logits before softmax activation.

Since softmax function is the soft function of $\text{onehot}(\arg \max(\cdot)))$, the  ${\bf l}_j$ can be approximated by 
\begin{equation}\label{eq20}
    {\bf 1}_j \approx {\bm p} = \text{softmax}({\bm l})  
\end{equation}
where ${\bm p}$ is the predicted probabilities.

Submitting the \eqref{eq20} to \eqref{eq18}, the answer accuracy can be approximated as
\begin{equation}\label{eq21}
    Acc \approx \frac{1}{{\left| B \right|}}\sum\limits_{ B} {\left\langle {{{\mathbf{1}}_i},{{{\bm p}}}} \right\rangle } = \frac{1}{{\left| B \right|}}\sum\limits_{ B}p(a)p(\hat a).
\end{equation}
where $p(a)$ is the real probability for label $i$ and  $p(\hat a)$ is the $i$-th  predicted probability at ${\bm p}$. 

Based on \eqref{eq21}, the loss function of answer accuracy can be designed as
\begin{equation}
    {\cal L}_{\tt Acc} = -\mathbb{E}\left[p(a) p(\hat a) \right].
\end{equation}

The derivation of ${\cal L}_{\tt Acc}$ for the parameters $\bm \varphi$ is
\begin{equation}\label{eq23}
    {\nabla _{\bm \varphi} }{\cal L}_{\tt Acc} = p(\hat a)\left( {1 - p(\hat a)} \right){\nabla _{\bm \varphi} }{\bm l}.
\end{equation}
From \eqref{eq23}, there exist two optimization directions when ${\nabla _{\bm \varphi} }{\cal L}_{\tt Acc} \to 0$, i.e., $p(\hat a) \to 0$ and $p(\hat a) \to 1$. However, $p(\hat a) \to 0$ causes worse prediction results and should avoid. In order to make the optimization stable, the ${\cal L}_{\tt Acc}$ should be refined. One refined loss function is the cross-entropy loss function given by
\begin{equation}
    {\cal L}_{\tt CE}= -\mathbb{E}\left[p(a) \log \left( p(\hat a) \right)\right].
\end{equation}

The derivation of ${\cal L}_{\tt CE}$ for the parameters $\bm \varphi$ is
\begin{equation}\label{eq25}
    {\nabla _{\bm \varphi} }{\cal L}_{\tt Acc} = \left( {1 - p(\hat a)} \right){\nabla _{\bm \varphi} }{\bm l}.
\end{equation}

Compared \eqref{eq23} and \eqref{eq25}, the derivation of ${\cal L}_{\tt CE}$ only has one correct optimization direction $p(\hat a) \to 1$, which is more stable during training. Therefore, the proposition \ref{prop1} is derived.

\section{Proof of Proposition 2}
For the classification task, the mutual information, $I\left(\tilde{\bm x}^c(k); a\right)$, can be expressed as
\begin{equation}\label{eq26}
     I\left(\tilde{\bm x}^c(k); a\right)=H(a) - H(a|\tilde{\bm x}^c(k)).
\end{equation}
where $H(a)$ is the entropy of the real label, $H(a|\tilde{\bm x}^c(k))$ is the conditional entropy. 

The cross-entropy between the real label and the predicted label given $\tilde{\bm x}^c(k)$ is 
\begin{equation}
    H(a; {\hat a}|\tilde{\bm x}^c(k)) = H(a|\tilde{\bm x}^c(k)) + D_{\text{ KL}}\left(a ||{\hat a}|\tilde{\bm x}^c(k) \right),
\end{equation}
where $D_{\text KL}\left(\cdot || \cdot \right)$ is the Kullback–Leibler divergence and is always non-negative. Therefore, we have the following inequality
\begin{equation}\label{eq28}
     H(a; {\hat a}|\tilde{\bm x}^c(k))  \geqslant  H(a|\tilde{\bm x}^c(k)),
\end{equation}

Submitting \eqref{eq28} into \eqref{eq26}, the lower bound of $I\left(\tilde{\bm x}^c(k); a\right)$ can be obtained 
\begin{equation}\label{eq29}
    I\left(\tilde{\bm x}^c(k); a\right)\geqslant H(a) -H(a; {\hat a}|\tilde{\bm x}^c(k)).
\end{equation}

From \eqref{eq29}, since $H(a)$ is constant, maximizing the $I\left(\tilde{\bm x}^c(k); a\right)$ can be approximated to minimizing the $H(a; {\hat a}|\tilde{\bm x}^c(k))$. The lower bound will be closer to $I\left(\tilde{\bm x}^c(k); a\right)$ when the model is trained. Therefore, the proposition \ref{prop-2} is derived.

\section{Proof of Proposition 3}
For the regression task, the mutual information, $I\left(\tilde{\bm x}^c(k); a\right)$, can be expressed as \eqref{eq26}.
\begin{lemma}
The conditional differential entropy yields a lower bound on the expected squared error of an estimator, for any random variable $X$, observation $Y$, and estimator $\hat X$, the following holds
\begin{equation}
    \mathbb{E}\left[\left(X- \hat X\left( Y \right) \right)^2 \right] \geqslant \frac{1}{2\pi e}e^{2H(X|Y)}.
\end{equation}
\end{lemma}

Applying the Lemma 1, the upper bound of conditional entropy, $H(a|\tilde{\bm x}^c(k))$, can be expressed as
\begin{equation}\label{eq31}
    H(a|\tilde{\bm x}^c(k)) < \mathbb{E}\left[ \ln\left| a - \hat a(\tilde{\bm x}^c(k))\right| \right]
\end{equation}
where $\hat a(\tilde{\bm x}^c(k))$ means the model outputs $\hat a$ with the $\tilde{\bm x}^c(k)$.

Submitting \eqref{eq31} into \eqref{eq26}, the lower bound of $I\left(\tilde{\bm x}^c(k); a\right)$ can be obtained
\begin{equation}\label{eq32}
     I\left(\tilde{\bm x}^c(k); a\right) > H(a) - \mathbb{E}\left[ \ln\left| a - \hat a\right| \right].
\end{equation}
From \eqref{eq32}, since $H(a)$ is constant, maximizing the $I\left(\tilde{\bm x}^c(k); a\right)$ can be approximated to minimizing the $\mathbb{E}\left[ \ln\left| a - \hat a\right| \right]$.  However, directly minimizing the $\mathbb{E}\left[ \ln\left| a - \hat a\right| \right]$ may cause the gradient explosion.

Given the derivation of $\ln\left| a - \hat a\right|$ for the parameters ${\bm \varphi}$,
\begin{equation}\label{eq33}
    {\nabla_{\bm \varphi} }\ln \left| {a - \hat a} \right| = \frac{1}{{\left| {a - \hat a} \right|}}{\nabla _{\bm \varphi} }\hat a.
\end{equation}
From \eqref{eq33}, when $\hat a \to a$, ${\nabla_{\bm \varphi} }\ln \left| {a - \hat a} \right| \to \infty$. In order to alleviate the gradient explosion, the approximation of $\ln \left| {a - \hat a} \right|$ is derived by applying the Taylor series expansion
\begin{equation}\label{eq34}
    \ln\left( \left| a - \hat a\right| - 1 + 1 \right) \approx \left|a - \hat a \right| - 1.
\end{equation}

The derivation of \eqref{eq34} for the parameters ${\bm \varphi}$ is
\begin{equation}\label{eq35}
    {\nabla_{\bm \varphi} } \left| {a - \hat a} \right| = {\nabla _{\bm \varphi} }\hat a.
\end{equation}

Compared \eqref{eq35} and \eqref{eq33}, the item, $\frac{1}{{\left| {a - \hat a} \right|}}$, is removed, therefore,  the gradient explosion is eliminated. Then, the lower bound of $I\left(\tilde{\bm x}^c(k); a\right)$ can be expressed as
\begin{equation}\label{eq36}
    I\left(\tilde{\bm x}^c(k); a\right)> H(a) - \mathbb{E}\left[ \left| a - \hat a\right| \right].
\end{equation}
From \eqref{eq36},  maximizing the $I\left(\tilde{\bm x}^c(k); a\right)$ can be approximated to minimizing the $\mathbb{E}\left[ \left| a - \hat a\right| \right]$. The lower bound will be closer to $I\left(\tilde{\bm x}^c(k); a\right)$ when the model is trained. Therefore, the proposition \ref{prop-3} is derived.



\ifCLASSOPTIONcaptionsoff
  \newpage
\fi

\bibliographystyle{IEEEtran}
\bibliography{IEEEabrv, reference.bib}

\end{document}